\documentstyle[preprint,prd,aps]{revtex}
\begin{document}
\draft
%\twocolumn[\hsize\textwidth\columnwidth\hsize\csname@twocolumnfalse\endcsname
\title{Meson propagators in spontaneously broken gauge theories}
\author{Hung Cheng~\cite{hun}} 
\address{Department of Mathematics, MIT, Cambridge, MA 02138, USA} 
\author{S. P. Li~\cite{li}}
\address{Institute of Physics, Academia Sinica, Taipei, Taiwan 115, R.O.C.}
\date{July, 2001}
\maketitle

\begin{abstract}
Contrary to common belief, the longitudinal vector meson and scalar meson 
propagators
have very different forms in spontaneously broken gauge  
theories.  We choose the Abelian-Higgs model as an example and show
that the longitudinal vector meson and scalar meson propagators 
have double poles  
by an extensive use of the Ward-Takahashi identities.  
\end{abstract}

%\vspace{2pc}]

Spontaneously broken gauge theories have been with us for a long
time.  A well-known example is the Electroweak Standard Model[1].   
Despite being studied for several decades, some of the  
peculiar features of spontaneously broken gauge theories 
have never been mentioned in the literature.  In this
paper, we will give a discussion of the vector and scalar meson
propagators of spontaneously broken gauge theories.  
Propagators in quantum field theories generally have simple poles.
This is indeed the case for the vector and scalar meson propagators
before symmetry is spontaneously broken.  Such a simple 
structure does not hold anymore after the symmetry is spontaneously
broken.  In the following, we will show this
by the extensive use of the Ward-Takahashi identities.  To make
our discussion simple, we will use the Abelian-Higgs model as an
example.  We will start with the unrenormalized Lagrangian and obtain all
our results in the unrenormalized form.  We will give a discussion on  
the renormalized case and the Electroweak Standard Model at the
end of the paper.

The Lagrangian for the Abelian-Higgs theory is

$$
L = - \frac{1}{4} F_{\mu\nu} F^{\mu\nu} + (D^\mu \phi)^{\dagger}
(D_\mu \phi) - V \,\, , 
\eqno(1)
$$
where
$$
F_{\mu\nu} = \partial_\mu A_\nu - \partial_\nu A_\mu \,\, ,
\,\, D_\mu \phi = (\partial_\mu + i g A_\mu) \phi \,\, , 
$$
$$
V = - \mu^2 \phi^{\dagger} \phi + \lambda (\phi^{\dagger} \phi)^2 \,\, , 
\eqno(2)
$$
and $g$, $\mu^2$ and $\lambda$ are the gauge coupling, the two-point 
and four-point scalar couplings respectively.  

The effective Lagrangian in the $\alpha$-gauge is
$$
L_{eff} = L + L_{gf} + L_{ghost} \,\, , 
\eqno(3)
$$
where 
$$
L_{gf} = - \frac{1}{2 \alpha} (\partial^\mu A_\mu -
\alpha \Lambda \phi_2)^2 \,\, ,
\eqno(4)
$$
and
$$
L_{ghost} = - i \partial_\mu \eta \partial^\mu \xi 
+ i \alpha \Lambda M \eta \xi + i \alpha g 
\Lambda H \eta \xi \,\, ,
\eqno(5)
$$
with
$$
\displaystyle
\phi = \frac{v + H + i \phi_2}{\sqrt{2}} \,\,\,\, 
{\text {and}} \,\,\,\, M = g v \,\, .
\eqno(6)
$$
The quantity $v$ is so defined that the vacuum expectation value
of $H$ is zero, i.e. 
$$
< H > = 0 \,\, .
\eqno(7)
$$
The variations of the fields under the BRS[2] transformations
are:
$$
\delta A_\mu = \partial_\mu \xi \,\, ,
\eqno(8a)
$$
$$
\delta \phi_2 = - g \xi (v + H) \,\, ,
\eqno(8b)
$$
$$
\delta H = g \xi \phi_2 \,\, ,
\eqno(8c)
$$
$$
\delta i \eta = \frac{1}{\alpha} \partial^\mu A_\mu - \Lambda \phi_2 \,\, ,
\eqno(8d)
$$
and 
$$
\delta \xi = 0 \,\, .
\eqno(8e)
$$
All fields and parameters are bare.

In the Abelian-Higgs model, a longitudinal $A$ meson may propagate into 
either a longitudinal $A$ or the $\phi_2$ scalar meson.  Thus, all 
of the following propagators
$$
\begin{array}{cc}
< 0 | T A_\mu(x) A_\nu(0) | 0 > \,\,\, ,& \,\, 
< 0 | T A_\mu(x) \phi_2(0) | 0 > \,\,\, , \\
< 0 | T \phi_2(x) A_\mu(0) | 0 > \,\,\, ,& \,\,
< 0 | T \phi_2(x) \phi_2(0) | 0 > \,\,\, ,\\
\end{array}
\eqno(9)
$$
are non-zero and together they form a $2 \times 2$ mixing matrix.  
We shall denote the Fourier transform of such a propagator with the
symbol $G$ and put 
$$
G^{AA}_{\mu\nu}(k^2) = D_{AA}^T(k^2) T_{\mu\nu} +  D_{AA}(k^2)
L_{\mu\nu} \,\, ,
\eqno(10a)
$$
$$
\displaystyle 
G^{A\phi_2}_\mu(k^2) = - G^{\phi_2 A}_\mu(k^2) = \frac{k^\mu}{k} 
D_{A\phi_2}(k^2) \,\, ,
\eqno(10b)
$$
$$
G^{\phi_2\phi_2}(k^2) = D_{\phi_2\phi_2}(k^2) \,\, ,
\eqno(10c)
$$
where
$$
\displaystyle 
T_{\mu\nu} = g_{\mu\nu} - \frac{k_\mu k_\nu}{k^2} \,\, , \,\,
L = \frac{k_\mu k_\nu}{k^2} \,\, .
$$ 
and $k = \sqrt{k^2}$.  
The scalar $\phi_2$ mixes the longitudinal $A$ but not with the
transverse $A$, as indicated by the factor $k^\mu$ in the expression
for $G^{A \phi_2}$.  Thus, in the $2 \times 2$ mixing matrix under
discussion, this factor $k^\mu$ will be replaced by $k$.
For the same reason, $L_{\mu\nu}$ can be replaced by unity.  The 
$2 \times 2$ mixing matrix is therefore equal to 
$$
\left[
\matrix{  D_{AA} & \displaystyle  D_{A\phi_2} \cr 
\displaystyle   D_{\phi_2 A} & D_{\phi_2 \phi_2} \cr }
\right] \,\, .
\eqno(11)
$$ 
We shall express the propagators by their 1PI amplitudes.

In a field theory without mixing, let the unperturbed propagator for
a particle be $i (k^2 - m^2)^{-1}$ with $m^2$ its unperturbed mass
squared.  If we call $\Pi$ to be its 1PI self-energy amplitude, then  
this propagator is the inverse of $-i (k^2 - m^2 - \Pi)$ or, 
$ -i (k^2 - \Gamma)$, where $\Gamma = m^2 + \Pi$.  Extending this
to our case, we can express the inverse of the 
mixing matrix as
$$
i \left[
\matrix{ \displaystyle \frac{k^2}{\alpha} - \Gamma_{AA}(k^2) 
& -i k (\Lambda - \Gamma_{A \phi_2}(k^2)) \cr  
i k (\Lambda - \Gamma_{A \phi_2}(k^2)) 
& \alpha \Lambda^2 - k^2 \Gamma_{\phi_2\phi_2}(k^2) \cr } 
\right] \,\, .
\eqno(12)
$$ 
where the unperturbed value of the $\Gamma$ functions are
$\Gamma_{AA} = M^2 \,\, , \,\, \Gamma_{A\phi_2} = M $  
and $\Gamma_{\phi_2\phi_2} = 1$.

The Ward-Takahashi identities for the propagators can now be
written down easily,
$$
\displaystyle
< T (\frac{1}{\alpha} \partial^\mu A_\mu(x) - \Lambda \phi_2(x))
(\frac{1}{\alpha} \partial^\nu A_\nu(x) - \Lambda \phi_2(x)) > = 0 \,\, ,
\eqno(13)
$$
$$
\displaystyle
< T (\frac{1}{\alpha} \partial^\mu A_\mu(x) - \Lambda \phi_2(x)) A^\nu(y) > 
= < T i \eta(x) \partial^\nu \xi(y) > \,\, ,
\eqno(14)
$$
and
$$
\displaystyle
< T (\frac{1}{\alpha} \partial^\mu A_\mu(x) - \Lambda \phi_2(x)) \phi_2(x) >
= - g < T i \eta(x) \xi(y) (v + H(y)) > \,\, .
\eqno(15)
$$
Equations (13)-(15) give 
$$
\displaystyle
\frac{k^2}{\alpha^2} D_{AA}(k^2) + \frac{2ik}{\alpha} \Lambda D_{A\phi_2}(k^2)
+ \Lambda^2 D_{\phi_2\phi_2}(k^2) = - \frac{i}{\alpha} \,\, , 
\eqno(16)
$$
$$
\displaystyle
- \frac{ik}{\alpha} D_{AA}(k^2) + \Lambda D_{A\phi_2}(k^2) 
= ik D_{\eta\xi}(k^2) \,\, , 
\eqno(17)
$$
and
$$
\displaystyle
- \frac{ik}{\alpha} D_{A\phi_2}(k^2) - \Lambda D_{\phi_2\phi_2}(k^2)
= - gv D_{\eta\xi}(k^2) - g D_{\eta\xi}(k^2) \Gamma_{\eta(\xi H)}(k,-k)
\,\, , 
\eqno(18)
$$ 
where $\Gamma_{\eta(\xi H)}$ is an abnormal amplitude with $\xi$ and $H$
joined together.   

Using (16), we get
$$
\Gamma_{AA}(k^2) \Gamma_{\phi_2\phi_2}(k^2) = (\Gamma_{A\phi_2}(k^2))^2 \,\, .
\eqno(19)
$$
which then gives
$$
\displaystyle
D_{AA}(k^2) = i\alpha \Gamma_{AA}(k^2) \frac{\alpha\Lambda^2 
- k^2\Gamma_{\phi_2\phi_2}(k^2)}{J^2(k^2)} \,\, ,
\eqno(20)
$$
$$
\displaystyle 
D_{\phi_2\phi_2}(k^2) = i\Gamma_{AA}(k^2) 
\frac{k^2 - \alpha\Gamma_{AA}(k^2)}{J^2(k^2)} \,\, ,
\eqno(21)
$$  
and
$$
\displaystyle 
D_{A\phi_2}(k^2) = - D_{\phi_2 A}(k^2) = - \alpha k \Gamma_{AA}(k^2)
\frac{\Lambda - \Gamma_{A\phi_2}(k^2)}{J^2(k^2)} \,\, ,
\eqno(22)
$$
where
$$
J(k^2) = \alpha \Lambda \Gamma_{AA}(k^2) - k^2 \Gamma_{A\phi_2}(k^2) \,\, .
\eqno(23)
$$
From (20)-(22), we get
$$
\displaystyle
k F_A(k^2) \equiv - \frac{ik}{\alpha} D_{AA}(k^2) - \Lambda D_{\phi_2 A}(k^2)
= \frac{k \Gamma_{A\phi_2}(k^2)}{J(k^2)} \,\, ,
\eqno(24)
$$
and
$$
\displaystyle 
F_{\phi_2}(k^2) \equiv - \frac{ik}{\alpha} D_{A\phi_2}(k^2) 
- \Lambda D_{\phi_2 \phi_2}(k^2) = i \frac{\Gamma_{AA}(k^2)}{J(k^2)} \,\, .
\eqno(25)
$$
 
One can see that the denominator in (20)-(22) is the square of $J$. 
Since the poles of these propagators come from the zeroes of $J$, and 
since $J$ is a linear superposition of 1PI self-energy amplitudes,  
which are analytic functions of $k^2$, the order of the poles of
the propagators are always even. 

It is known that propagators in quantum field theories generally have
simple poles.  Contrary to common belief, one can demonstrate that
this is often not true when two fields mix under the auspices of
the Ward-Takahashi identities.  In the case studied here, the  
longitudinal A and the scalar meson have the same unperturbed mass to
begin with.  The poles for these fields are both at $k^2 = \alpha \Lambda M$.
As the interactions are turned on and the propagators form a mixing
matrix, the positions of the poles change but the Ward-Takahashi 
identities force them to remain to be the same point.  Thus the
two simple poles merge to form a double pole.  

On the other hand, the ghost propagator does not have the above 
double pole structure.  Let the Fourier transform of 
$< 0 | T i \eta(x) \xi(0) | 0 >$ be denoted as
$$
D_{\eta\xi}(k^2) \,\, . 
\eqno(26)
$$
From (17) and (24), we get
$$
\displaystyle
D_{\eta \xi}(k^2) = - \frac{i \Gamma_{A\phi_2}(k^2)}{J(k^2)} \,\, .
\eqno(27)
$$ 

The only gauge that still bears a simple pole for 
the meson propagators is the Landau gauge.  This is done by setting 
$\alpha$ to go to zero.  In this case, the only propagator that is
nonvanishing in the mixing matrix (11) is $G^{\phi_2\phi_2}$ which
takes the value
$$
\displaystyle 
G^{\phi_2\phi_2}(k^2) = \frac{i}{k^2 \Gamma_{\phi_2\phi_2}(k^2)} \,\, .
\eqno(28)
$$
The ghost propagator decouples from the theory and takes the form
$$
\displaystyle 
D^{\eta \xi}(k^2) = \frac{i}{k^2} \,\, .
\eqno(29)
$$   
Therefore, both the propagators of (28) and (29) are of simple poles. 

If one adds fermion fields to the theory, the Lagrangian given in (1) 
will have additional terms.  When the fermions are chiral fermions, parity
is not conserved.  If one assumes that the expanded Lagrangian still
possess the same BRS invariance and that the fermion fields are 
introduced in such a way that there is no anomaly, the Ward-Takahashi
identities given above still holds, and thus the vector and 
scalar meson propagators
will still have the same form. 

The above discussion can be extended easily to the Electroweak 
Standard Model.  There are now four vector mesons to
begin with.  The positively charged W mesons will mix with the positively
charged scalar to form a $2 \times 2$ mixing matrix.  Similarly for
the negatively charged mesons.  The other two
mesons, the $Z$ meson and the photon (or, in terms of the original
$SU(2)$ and $U(1)$ gauge fields, the neutral $W$ meson and the $B$
meson) will mix with the scalar partner of the Higgs and form a 
$3 \times 3$ mixing matrix.  One can carry out a similar but somewhat 
complicated derivation to demonstrate that the double pole structures 
again appear in the vector meson and scalar meson
propagators[3].  

In summary, we have shown that propagators in spontaneous broken
gauge theories can have different pole structures once the symmetry
is broken spontaneously.  This is done by an extensive use of the
Ward-Takahashi identities.  The fact that the propagators have double 
poles would make the
renormalization of spontaneously broken gauge theories, such as the
Electroweak Standard Model more involved 
in covariant $\alpha$ gauges, let alone the Landau gauge.  A study
of this has been carried out recently[3] and will be presented elsewhere.

\end{document}